\documentclass[aps,prx, reprint, superscriptaddress, footinbib]{revtex4-1}
\usepackage[T1]{fontenc}
\usepackage[utf8]{inputenc}
\usepackage{graphicx}
\usepackage{bm}
\usepackage{color}
\usepackage{verbatim}
\usepackage{soul}
\usepackage{amsmath}

\begin{document}

\title{Observation of Brillouin optomechanical strong coupling \\with an 11 GHz mechanical mode}

\author{G.~Enzian}
\email{georg.enzian@physics.ox.ac.uk}
\affiliation{Clarendon Laboratory, Department of Physics, University of Oxford, OX1 3PU, United Kingdom}

\author{M.~Szczykulska}
\affiliation{Clarendon Laboratory, Department of Physics, University of Oxford, OX1 3PU, United Kingdom}

\author{J.~Silver}
\affiliation{National Physical Laboratory (NPL), Teddington, TW11 0LW, United Kingdom}

\author{L.~Del~Bino}
\affiliation{National Physical Laboratory (NPL), Teddington, TW11 0LW, United Kingdom}

\author{S.~Zhang}
\affiliation{National Physical Laboratory (NPL), Teddington, TW11 0LW, United Kingdom}

\author{I. A. Walmsley}
\affiliation{Clarendon Laboratory, Department of Physics, University of Oxford, OX1 3PU, United Kingdom}

\author{P. Del'Haye}
\affiliation{National Physical Laboratory (NPL), Teddington, TW11 0LW, United Kingdom}

\author{M. R. Vanner}
\email{m.vanner@imperial.ac.uk}
\affiliation{Clarendon Laboratory, Department of Physics, University of Oxford, OX1 3PU, United Kingdom}
\affiliation{QOLS, Blackett Laboratory, Imperial College London, London SW7 2BW, United Kingdom}

%
%\affil[1]{Clarendon Laboratory, Department of Physics, University of Oxford, OX1 3PU, United Kingdom}
%\affil[2]{National Physical Laboratory (NPL), Teddington, TW11 0LW, United Kingdom}
%\affil[3]{QOLS, Blackett Laboratory, Imperial College London, London SW7 2BW, United Kingdom}
%\affil[*]{Corresponding authors: georg.enzian@physics.ox.ac.uk (G.E.); m.vanner@imperial.ac.uk (M.R.V.)}

%
%\author[1, 2]{\small Erwin T. Lau}
%\author[3, 5]{\small Massimo Gaspari}
%\author[1, 2, 4]{\small Daisuke Nagai}
%\author[1, 2, 4]{\small Paolo Coppi}
%
%\affil[1]{\footnotesize Department of Physics, Yale University, New Haven, CT 06520, USA}
%\affil[2]{\footnotesize Yale Center for Astronomy and Astrophysics, Yale University, New Haven, CT 06520, USA}
%\affil[3]{\footnotesize Department of Astrophysical Sciences, Princeton University, 4 Ivy Lane, Princeton, NJ 08544-1001 USA}
%\affil[4]{\footnotesize Department of Astronomy, Yale University, New Haven, CT 06520, USA}
%\affil[5]{\footnotesize Einstein and Spitzer Fellow}

%\setcounter{Maxaffil}{0}

%\ociscodes{(140.3490) Lasers, distributed feedback; (060.2420) Fibers, polarization-maintaining; (060.3735) Fiber Bragg gratings.}

% To be edited by editor
% \doi{\url{http://dx.doi.org/10.1364/optica.XX.XXXXXX}}

\begin{abstract}
Achieving cavity-optomechanical strong coupling with high-frequency phonons provides a rich avenue for quantum technology development including quantum state-transfer, memory, and transduction, as well as enabling several fundamental studies of macroscopic phononic degrees-of-freedom. Reaching such coupling with GHz mechanical modes however has proved challenging, with a prominent hindrance being material- and surface-induced-optical absorption in many materials. Here, we circumvent these challenges and report the observation of optomechanical strong coupling to a high frequency (11 GHz) mechanical mode of a fused-silica whispering-gallery microresonator via the electrostrictive Brillouin interaction. Using an optical heterodyne detection scheme, the anti-Stokes light backscattered from the resonator is measured and normal-mode splitting and an avoided crossing are observed in the recorded spectra, providing unambiguous signatures of strong coupling. The optomechanical coupling rate reaches values as high as $G/2\pi = 39 \ \text{MHz}$ through the use of an auxiliary pump resonance, where the coupling dominates both the optical ($\kappa/2\pi = 3 \ \text{MHz}$) and the mechanical ($\gamma_\text{m}/2\pi = 21 \ \text{MHz}$) amplitude decay rates. Our findings provide a promising new approach for optical quantum control using light and sound.
\end{abstract}

\maketitle

\section{Introduction}
Since the 1920s photon-phonon Brillouin scattering \cite{Brillouin:22, Mandelstam:26, Raman:28} has been a subject of intense and diverse study. This nonlinear optical phenomenon has been observed in numerous physical systems including bulk crystals \cite{Chiao:64}, optical fibres \cite{Ippen:72, Agrawal:13}, integrated devices such as silicon photonic waveguides \cite{VanLaer:15}, silica micro-sphere resonators \cite{Tomes:09}, and bulk crystalline resonators \cite{Renninger:18}. The interaction is now receiving a resurge of interest and its potential to contribute to both classical- and quantum-information-processing applications has recently been identified. Prominent example applications of Brillouin scattering include optical delay and memory \cite{Zhu:07, Merklein:17}, coherent waveguide interfacing and filtering \cite{Merklein:17, Marpaung:15, Shin:15}, non-reciprocity \cite{Dong:15, Shen:16}, as well as switching, pulse shaping, and other optical technologies such as fibre Brillouin lasers and amplifiers \cite{Garmire:17}. Moreover, Brillouin scattering provides a bridge between light and sound and offers an attractive path to coherently connect the microwave and optical domains.

The field of Brillouin scattering is now merging with the rapidly growing field of cavity quantum optomechanics. This merger offers new opportunities to control phononic degrees-of-freedom at the quantum level to develop new applications, such as weak-force sensors, and to study the fundamentals of quantum physics at a macroscopic scale. In quantum optomechanics, radiation-pressure is one of the most commonly employed interactions \cite{Aspelmeyer:14}, which should be contrasted to electrostriction and photoelasticity which are central to the physics of Brillouin scattering. It is of vital importance for many optomechanical protocols that a large coupling rate between the electromagnetic and mechanical degrees-of-freedom be achieved. As in other facets of quantum optics, such as cavity- \cite{Thompson:92, Raimond:01} and circuit-quantum electrodynamics \cite{Wallraff:04}, it is important to achieve a coupling rate that exceeds the decay rates present in the system. This strong coupling regime has been theoretically studied for an optical cavity field coupling to a mechanical oscillator \cite{Marquardt:07, Dobrindt:08} and has now been observed in optomechanical \cite{Groblacher:09} and electromechanical \cite{Teufel:11} systems where the characteristic Rabi-like splitting of the mechanical power-spectral density was demonstrated. It is also important in optomechanics that the coupling rate exceeds the mechanical decoherence rate in order to provide coherent control of the mechanical degree-of-freedom. This coherent coupling regime has also been observed in both optomechanical \cite{Verhagen:12} and electromechanical systems \cite{Teufel:11b} and enables the exciting prospect of coherent quantum state transfer between light and mechanical motion.

Brillouin interactions \cite{Wolff:15, Sipe:16, VanLaer:16, Zoubi:16, Huy:16, Rakich:18}, have been primarily studied in the stimulated regime that gives a large Stokes-scattering signal, however, very recently anti-Stokes scattering has been gaining more attention which is relevant for quantum control applications in optomechanics. For instance, laser cooling of a MHz frequency mechanical mode was performed using forward Brillouin scattering \cite{Bahl:12}, similar forward Brillouin scattering was used to suppress Rayleigh scattering \cite{Kim:18}, thermal anti-Stokes scattering from a silicon waveguide was very recently observed \cite{VanLaer:17}, and ultra-long-lived high-frequency phonon modes have been identified in bulk crystalline resonators \cite{Renninger:18}. Brillouin optomechanics operating in the back-scattering regime unites several favourable properties. Notably, high mechanical frequencies (> 10 GHz) enable low thermal occupations to be reached by standard commercial cryogenics; high separability between the optical pump and the scattered optical signal can be more easily achieved owing to the large frequency separation and backward scattering direction; and high-bandwidth, multi-wavelength operations can be performed that enable advanced quantum-information processing capabilities. Observing strong coupling for GHz mechanical modes has eluded the optomechanics and Brillouin scattering communities thus far and entering this regime would enable these united favourable properties to be exploited for optomechanical quantum technology development and advancing the forefront of fundamental quantum science.

Here we report the observation of strong coupling between an optical whispering gallery mode of a fused silica microresonator and an 11 GHz mechanical travelling wave via Brillouin anti-Stokes scattering. This is achieved with a continuous-wave pump at 1550 nm that is resonantly enhanced by an auxiliary cavity mode. The pump field counter-propagates with the mechanical mode and scatters from thermal phonons to create a backward-propagating anti-Stokes optical signal, while Stokes scattering (which can lead to stimulated-Brillouin scattering) is not resonant with the cavity response and thus strongly suppressed. The anti-Stokes signal is then measured using heterodyne detection. As silica has an extremely low optical propagation loss a strong pump field can be utilized to achieve a coupling rate that far exceeds both the optical and the mechanical damping rates. Normal-mode splitting and an avoided crossing are observed in the heterodyne spectra providing unambiguous experimental signatures of strong coupling.

\section{Results}
Brillouin scattering is a three-wave mixing process where an optical field interacts with a mechanical travelling wave and generates a frequency shifted optical signal. There are two sides to this light-matter interaction: electrostriction, where the electric field of light influences the mechanical wave, and photoelasticity, where the mechanical wave modifies the light field. Momentum and energy conservation allow two types of scattering processes: Stokes scattering, where the frequency of the light is downshifted giving rise to optical gain and is commonly observed in the form of stimulated Brillouin scattering or Brillouin lasing; and anti-Stokes scattering, where the light is upshifted giving rise to a mechanical damping mechanism. The phase-matching conditions for the anti-Stokes process utilized in this work are shown in Fig. 1A. Here, an optical pump field interacts with a counter-propagating mechanical wave generating a back-scattered optical anti-Stokes field. Since the pump and anti-Stokes fields are of a similar frequency, the wavevectors in this back-scatter process are related via $|k_\text{m}| \approx 2 |k_\text{p}|\approx 2 |k_\text{aS}|$, where the subscripts m, p, and aS refer to the mechanical, optical pump, and anti-Stokes modes, respectively, see the supplementary material.

\subsection{Experimental Setup}
We used an optical microresonator that supports two optical whispering gallery modes spaced by approximately the mechanical frequency, see Fig. 1B. A pump laser drives the lower frequency auxiliary cavity mode (with small detuning $\delta_\text{p}$) to generate a large intracavity optical field. This field interacts with the mechanical mode and anti-Stokes light is resonantly scattered into the higher frequency optical mode. As there are only two optical modes that participate, the symmetry between Stokes and anti-Stokes scattering is broken and the Stokes scattering is strongly suppressed. We label the angular frequency mismatch between the mechanical frequency $\omega_\text{m}$ and the difference between the anti-Stokes resonance frequency $\omega_\text{aS}$ and the pump laser frequency $\omega_\text{L}$ by $\Delta = \omega_\text{aS} - \omega_\text{L} - \omega_\text{m}$. Note that the Brillouin frequency does not correspond to the free-spectral-range of the microresonator. Rather, the two optical modes with the desired frequency spacing are achieved by using different spatial modes that provide significant overlap with the mechanical traveling wave.
\begin{figure*}[htbp]
\centering
\includegraphics[width=\textwidth]{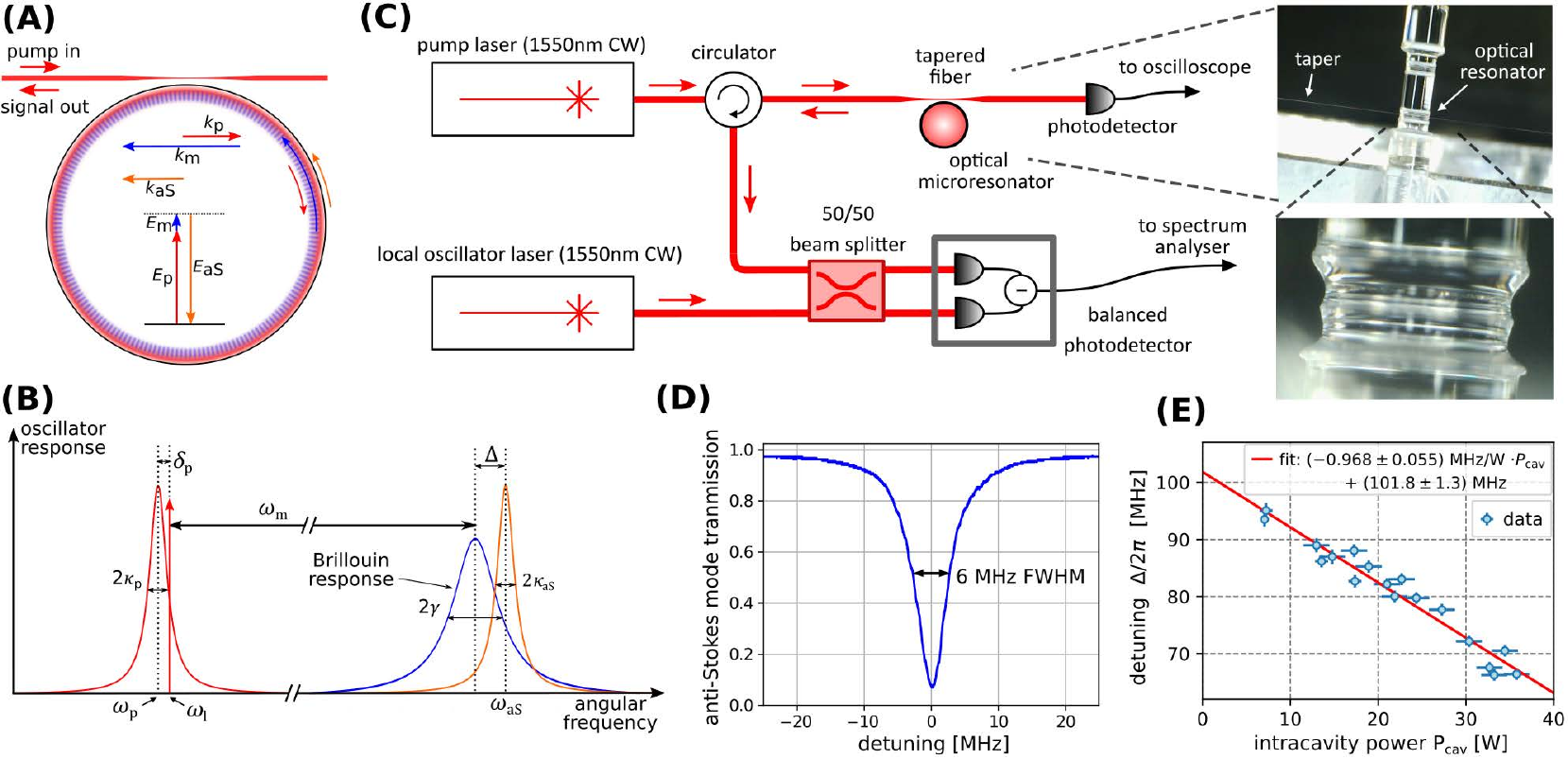}
\caption{\textbf{Experimental platform and setup.} (A) Wavevector and energy conservation in Brillouin anti-Stokes scattering. (B) Resonance structure of our Brillouin optomechanical system. Optical (red, orange) and mechanical (blue) modes with their frequencies and decay rates shown. (C) Experimental schematic (left) and optical microscope images of a 700 $\mu$m fused-silica microrod whispering gallery resonator (right). The microresonator is coupled to a tapered optical fiber and is driven by a continuous-wave pump laser. Frequency upconverted light backscattered from the resonator is separated by a circulator and measured using optical heterodyne detection and is recorded with a spectrum analyzer. (D) Taper transmission as the laser frequency is scanned across the anti-Stokes cavity resonance. (E) Observed detuning vs intracavity pump power due to optical nonlinearities.}
\label{fig:false-color}
\end{figure*}

The Brillouin frequency shift in bulk silica at 1550 nm is 10.7 GHz, which is obtained from the simple relation $\omega_\text{m} \approx 2 \omega v/(c/n)$. Here, $\omega_\text{m}$ is the Brillouin angular frequency shift, $v$ is the speed of sound in silica, $c$ is the speed of light in vacuum, $n$ is the refractive index, and $\omega$ is the optical angular frequency. To measure the Brillouin frequency shift we pumped the higher frequency mode of the optical mode pair and observed Brillouin lasing \cite{Tomes:09, Grudinin:09, Sturman:15} using an optical spectrum analyzer. We observed $\omega_\text{m}/(2 \pi) = (11.01 \pm 0.09) \ \text{GHz}$, which is consistent with previous stimulated Brillouin scattering measurements in silica microresonators \cite{Tomes:09} and indicates that electrostriction is the dominant coupling mechanism. The mechanical frequency in the anti-Stokes experiments discussed below must lie very close to this value. We would also like to note here that, unlike in conventional optomechanics, the Brillouin frequency shift has very little dependence on both the optical power and temperature providing a robust and convenient platform.

A schematic of our telecom-fiber-based experimental setup is shown in Fig. 1C. We utilize a fused silica micro-rod-resonator \cite{DelHaye:13} (diameter: $700 \ \mu\text{m}$, lateral radius of curvature: $\sim 40 \ \mu\text{m}$, free-spectral-range: 90 GHz) evanescently coupled to a tapered optical fiber. The lower frequency cavity resonance of the pair is driven by a continuous-wave pump laser and a thermal lock \cite{Carmon:04} is used, which stabilizes the resonance to the pump laser. The frequency upshifted light backscattered in the resonator is coupled out via the tapered fiber, separated from the pump light by an optical circulator, and mixed with a local oscillator on a 50:50 fiber-beam splitter. We then observe this signal with a balanced detector, implementing optical heterodyne detection with a local oscillator frequency offset of approximately 200~MHz. Heterodyne detection provides a large signal-to-electronic noise ratio and with the frequency offset allows the shape of the spectra to be easily observed. The heterodyne spectra are recorded using an electrical spectrum analyzer and the pump power is varied to characterize the Brillouin optomechanical strong coupling features. 

The optical damping rates are obtained from separate transmission spectra measurements. Fig. 1D shows the cavity mode that enhances the anti-Stokes signal, which has an amplitude decay rate of $\kappa_\text{aS}/2\pi = 3.0 \ \text{MHz}$. (We will use the convention of amplitude decay rates throughout the article.) Similarly, the pump mode was measured to have an amplitude decay rate of 3.5 MHz. As the mechanical frequency is orders of magnitude larger than the damping rate of the anti-Stokes optical mode, the experiment lies deeply within the resolved sideband regime, i.e. $\omega_\text{m} \gg \kappa_\text{aS}$, which strongly suppresses the Stokes scattering. These optical damping rates have an intrinsic contribution for which the major loss mechanisms have been identified \cite{Gorodetsky:96} and an external contribution due to the tapered optical fiber coupling (see supplementary materials).

By fitting to our heterodyne spectra, the mechanical amplitude decay rate was estimated to be $\gamma_\text{m}/2\pi = (20.9 \pm 1.6) \ \text{MHz}$. This value is similar to previous room-temperature in-fiber and bulk silica measurements at 1550 nm \cite{Boyd:08, Nikles:97}. We would like to highlight that it has been previously observed that the mechanical damping in such materials is significantly reduced when operating at low temperature, reducing by an order of magnitude at approximately 4 K \cite{Vacher:80, LeFloch:03}.

In our experiment, we observe that the detuning $\Delta$ decreases linearly with increasing intracavity pump power, see Fig. 1E. We attribute this detuning change to the optical Kerr effect and a possible contribution from the cavity-mode-dependent thermo-refractive effect, which can both cause pump-power-dependent relative frequency shifts between the two optical cavity modes. Concentrating on the former, self- and cross-phase modulation cause power-dependent shifts to the cavity resonances depending on the mode overlap and we can model the dependence of the detuning on the intracavity power by $\Delta \simeq \Delta_0 - P_\text{cav} \omega n_2/(n A')$ . Here, $\Delta$ is the detuning, $\Delta_0$ is the initial (low power) detuning, $\omega$ is the laser angular frequency, $n_2$ is the nonlinear refractive index, $n$ is the refractive index, $A'$ depends on the difference between the self- and cross-phase modulation terms and has dimensions of an area (see supplementary material), and $P_\text{cav}$ is the intracavity pump power. This purely optical mode overlap is different from the Brillouin optomechanical mode overlap, in that it involves only two (optical) modes, whereas the latter involves a triple overlap integral of one acoustic and two optical modes (see supplementary material). From the fit shown in Fig. 1E we observe a linear shift of approximately $1 \ \text{MHz} \ \text{W}^{-1}$ of intracavity power. The measurements in Fig. 1E were performed at low power in order to avoid nonlinear loss mechanisms, such as four-wave mixing parametric oscillation. This detuning measurement was also used as a calibration to determine the intracavity pump power in addition to the transmission measurement described in the methods section below. We would like to remark that it may be practically possible to engineer a cavity that eliminates the pump-power dependence of the detuning. This could be achieved by exploiting both the self- and cross-phase modulation and using an optical mode structure with an overlap such that the two modes have the same frequency shift as the pump power changes.

\subsection{Model}
The Brillouin interaction may be described by a simplified Hamiltonian that couples the two optical modes via the high-frequency mechanical oscillation. Since we are coherently driving the optical pump mode, we approximate its associated field operator by a classical amplitude, which acts to enhance the optomechanical coupling strength. In a frame rotating with the two optical frequencies, our simplified phenomenological model has the Hamiltonian
\begin{equation*}
\frac{H}{\hbar} = G \left( a_\text{aS}^\dag b + a_\text{aS} b^\dag \right) - \Delta b^\dag b \, .
\end{equation*}
Here, $a_\text{aS}$ and $b$ are the optical anti-Stokes mode and mechanical field operators, respectively, and $G= g_0 |\alpha| \propto \sqrt{P_\text{cav}}$ is the intracavity-pump-enhanced optomechanical coupling strength. Starting from this Hamiltonian, we compute the system dynamics using quantum Langevin equations. We then utilize optical input-output theory and compute the noise power spectral density of a rotating quadrature of the anti-Stokes field to describe the spectra observed with our heterodyne detection measurements (see supplementary material). It is important to note here that the present experiment operates in a regime where the acoustic density of states is a quasi-continuum, as the damping rate of each mechanical eigenfrequency component is larger than the mechanical free-spectral range. The mechanical field operator $b$ in the simplified model above then describes the linear combination of mechanical eigenfrequency components that contribute to the phase-matching. To model the system in this way, we assume that each mechanical frequency component couples equally and, as the optical linewidth is much smaller than the mechanical decay rate, this selects a narrow range of mechanical frequency contributions.

Consistent with our experimental observations detailed in the following section, our model indicates that, for zero detuning, the system undergoes normal-mode-splitting when $G > |\kappa_\text{aS} - \gamma_\text{m}|/2$. We would like to clarify that satisfying this condition does not necessarily demonstrate that strong coupling has been achieved as the two peaks in the spectra may not be clearly resolved. The conditions of strong coupling are met, when the coupling strength $G$ becomes larger than the effective damping rates of the hybrid optical-mechanical modes, i.e. $G > (\kappa_\text{aS} + \gamma_\text{m})/2$ (see supplementary materials). Under these conditions, a clearly separated avoided crossing may be observed in the spectra, which is an unambiguous signature of strong coupling.

\begin{figure}[htbp]
\centering
\includegraphics[width=0.9\linewidth]{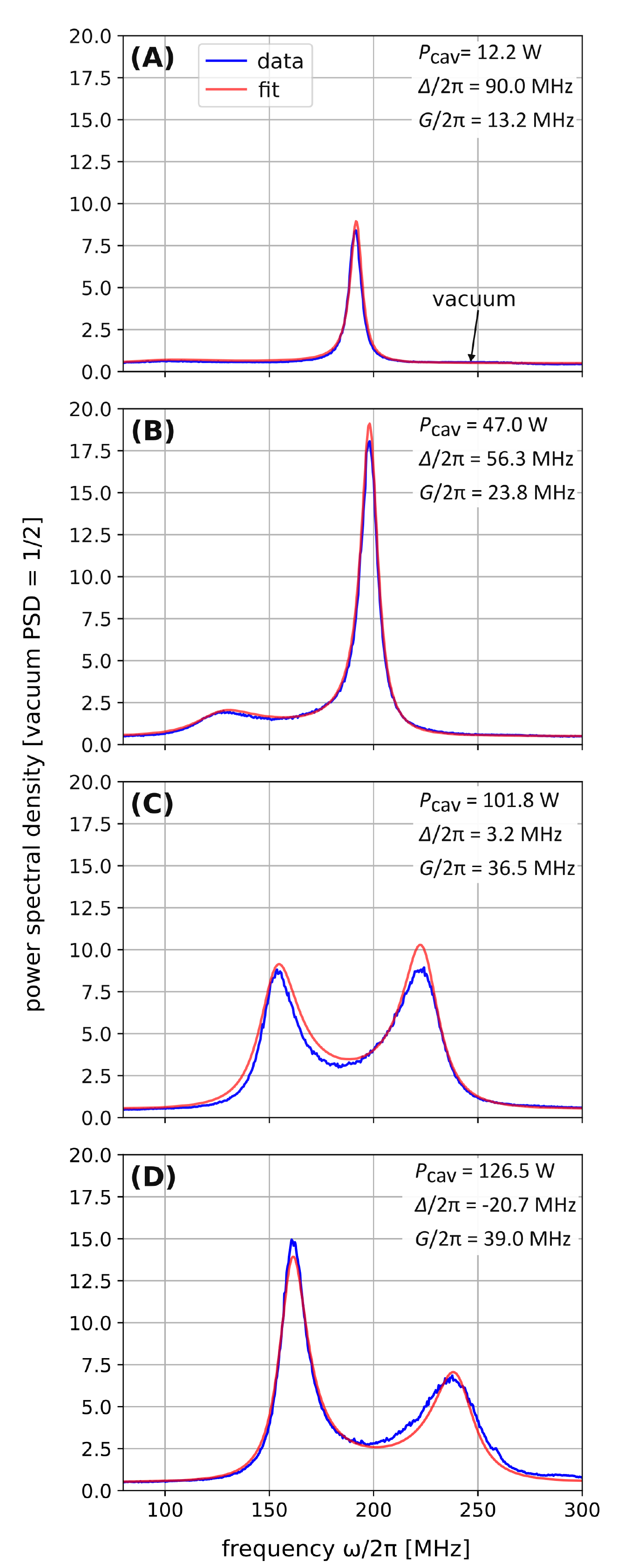}
\caption{\textbf{Observed Brillouin optomechanical anti-Stokes spectra.} Power spectral densities of the Brillouin anti-Stokes heterodyne signal (blue) with theoretical fits (red). The spectra are normalized such that a vacuum input corresponds to a value of 1/2. The heterodyne frequency is 190 MHz. As the intracavity pump power is increased from (A) through to (D), the detuning decreases, and the coupling rate increases. As the detuning goes through zero, an avoided crossing and normal mode splitting are observed, see plot (C).}
\label{fig:fig2}
\end{figure}

\subsection{Observation of normal-mode splitting and avoided crossing}
To characterize our Brillouin optomechanical system, we record the heterodyne spectra for a wide range of intracavity pump powers. As $P_\text{cav}$ increases, the optomechanical coupling rate increases in proportion to $\sqrt{P_\text{cav}}$, and simultaneously, the detuning $\Delta$ changes linearly with $P_\text{cav}$. We then record and fit the heterodyne spectra to obtain estimates of the experimental parameters, aside from the two optical decay rates measured previously, and excellent agreement between our model and the data is found. We would also like to highlight here that the optical resonator does not exhibit optical mode splitting through backscattering from imperfections of the material \cite{Kippenberg:02}, which enabled us to more easily confirm that the signals observed originate from the Brillouin optomechanical interaction. We plot and analyze the results of this work in terms of the intracavity pump power instead of the input pump power, so that the dependence on the pump detuning $\delta_\textrm{p}$ and taper coupling conditions is removed. This way also provides greater convenience, as the optomechanical coupling rate $G$ and detuning $\Delta$ directly depend on $P_\text{cav}$. At close to critical coupling, the intracavity power is approximately the input power multiplied by $\mathcal{F}/\pi \approx 4000$, where $\mathcal{F}$ is the cavity finesse.

In Fig. 2 a subset of the observed heterodyne spectra with theoretical fits is plotted. These plots show typical observed spectra, where the detuning varies from large and positive, through zero, to negative, whilst the opto-mechanical coupling rate increases. The spectra comprise a double peak structure on top of a flat optical vacuum background, where both the widths and center frequencies of the peaks change as the intracavity power changes. For low pump power and large positive detuning, the spectrum contains mainly a single narrow peak (Fig. 2A). As the pump power is increased the strength of the signal grows and a second side peak becomes more pronounced (Fig. 2B). As the power is further increased (Fig. 2C), two well separated approximately symmetric peaks are observed. At this power, the coupling rate dominates over all damping rates and the detuning in the system. For our particular physical implementation, the detuning passes through zero at this point and the heights and widths of the two peaks observed are approximately equal. The two peaks in the spectrum correspond to the in-phase and out-of-phase hybrid optical-mechanical modes, being the eigenstates of the system in the strong coupling regime. The peaks are spaced by $2G$, and their widths are given by the hybrid optical-mechanical damping rates. As the pump power is yet further increased, the peak to the left now becomes stronger and narrower compared to the peak on the right (Fig. 2D), and the separation between the two peaks further increases.

Fig. 3A plots the observed coupling rate with intracavity pump power for our complete set of measurements. The data fits very well to the model and the predicted scaling $G = g_0 |\alpha| \propto \sqrt{P_\text{cav}}$ is observed. From a fit to this data we observe that the coupling rate increases by $(3.605 \pm 0.016) \ \text{MHz} \ \text{W}^{-1/2}$ of intracavity pump power. To aid comparison we have overlaid the mechanical amplitude decay rate, the optical amplitude decay rate, and the hybrid-mode damping rate $(\kappa_\text{aS} + \gamma_\text{m})/2$, on this plot. It is seen that the coupling rate surpasses the hybrid-mode damping rate at an intracavity power of less than 10 W, corresponding to a very low input-pump power for these silica systems of only 2.5 mW. With increasing pump power, we can go deeply into the strong coupling regime, achieving a very high coupling rate of approximately 40 MHz, which exceeds the mean of the decay rates by a factor of more than 3.25. Using this fit result for $G$ and knowledge of the resonator geometry, we estimate the underlying Brillouin optomechanical coupling rate to be $g_0/2\pi = (396.5 \pm 1.8) \ \text{Hz}$, which is consistent with previous theoretical related work on Brillouin Stokes scattering \cite{Wolff:15, Sipe:16, VanLaer:16, Zoubi:16, Huy:16}.

\begin{figure*}[htbp]
\centering
\includegraphics[width=0.89\textwidth]{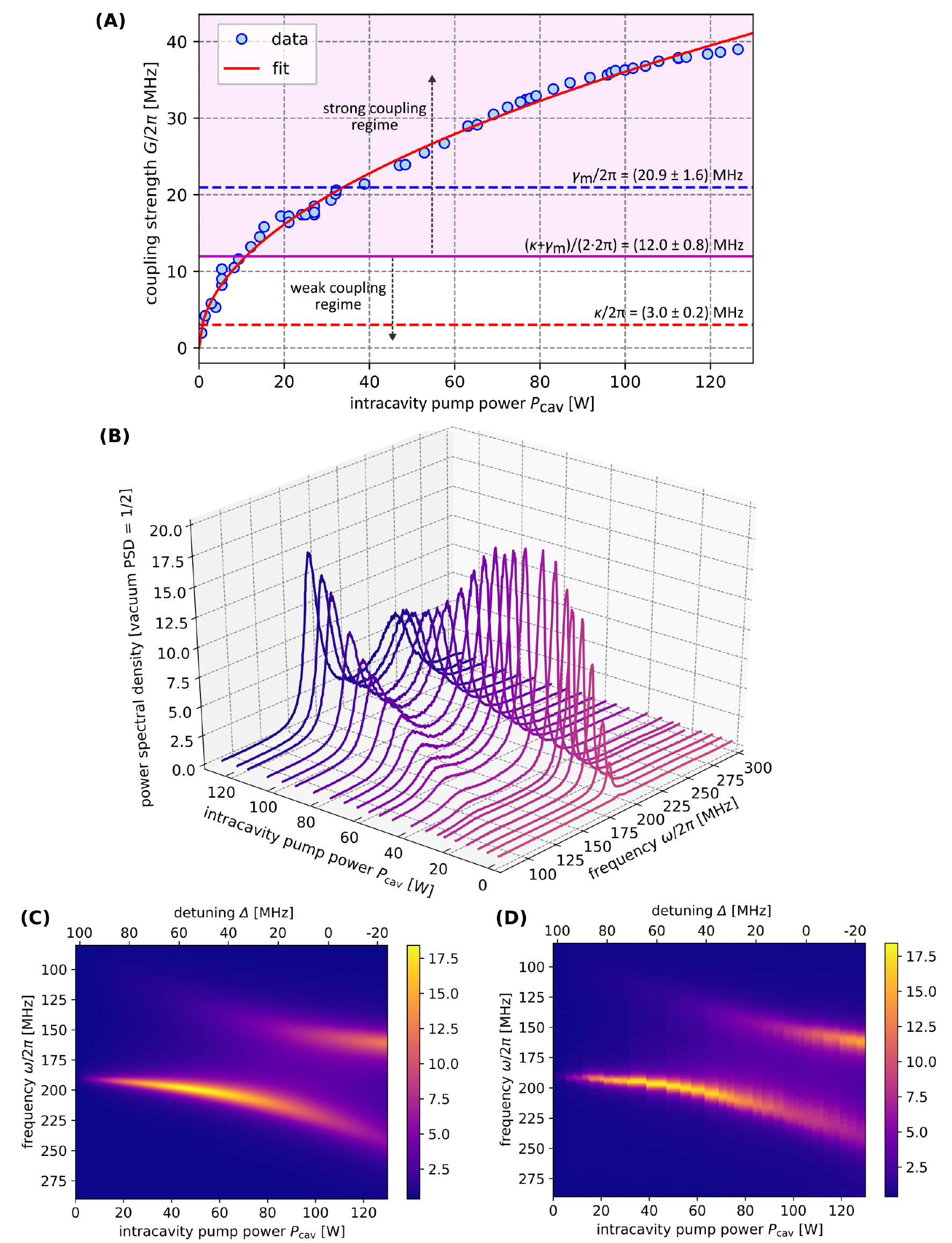}
\caption{\textbf{Brillouin optomechanical strong coupling showing normal mode splitting and an avoided crossing.} (A) Observed optomechanical coupling strength, as determined from the measured spectra, plotted with the intracavity pump power. The expected square-root scaling with the intracavity power is observed and the input-pump powers used were up to 30~mW. The coupling rate achieved far exceeds the damping rates of the system allowing us to go deeply into the strong coupling regime (shaded purple). (B) Experimentally observed spectra of the anti-Stokes scattered light with varying intracavity pump power. As the power increases, the point of zero detuning is crossed, where an avoided crossing is clearly observed. Theoretically predicted spectra (C), and observed spectra (D), with intracavity pump power. Note the excellent agreement between theory, which includes the power dependent-detuning, and experiment.}
\label{fig:fig3}
\end{figure*}

In Fig. 3B the observed evolution of the heterodyne spectrum with varying intracavity power is shown. As the intracavity power increases, a second lower frequency peak appears and grows, comes closer in frequency to the first peak and then the separation increases as the optomechanical coupling rate further increases. This is an avoided crossing, which is an unambiguous signature of strong coupling. In contrast to the more common avoided crossing plots, where the system eigenfrequencies are plotted against the detuning, here, we plot against the intracavity power, which acts as a proxy for the detuning because the detuning changes linearly with the intracavity power. Our theoretical model for the heterodyne spectrum with intracavity power is plotted in Fig. 3C for our experimental parameters. This model accounts for both the intracavity power dependent coupling and detuning, and we find excellent agreement with our observations plotted in Fig. 3D for comparison. An avoided crossing can be confirmed from our data (Fig. 2, Fig. 3B \& D) by noting that at the point where $\Delta \approx 0$ is reached ($G/2\pi \approx 36 \ \text{MHz} \ , \ \ P_\text{cav} \approx 100 \ \text{W}$) a splitting is observed in the spectrum. Such a splitting would not be present at $\Delta \approx 0$ if an avoided crossing were not present. For convenience, a list of the main experimental parameters is given in Table 1.

\begin{table}[htbp]
\centering
\caption{\bf Microresonator parameters achieving Brillouin optomechanical strong coupling.}
\begin{tabular}{|c|c|}
\hline
microresonator diameter & $700 \ \mu\text{m}$ \\
\hline
$\omega_\text{aS}/2\pi \approx \omega_\text{p}/2\pi$ & $193 \ \text{THz} \ (\lambda \sim 1550 \ \text{nm})$ \\
\hline
$\kappa/2\pi$ & $3.0 \ \text{MHz}$ \\
\hline
$\omega_\text{m}/2\pi$ & $11.0 \ \text{GHz}$ \\
\hline
$\gamma_\text{m}/2\pi$ & $20.9 \ \text{MHz}$ \\
\hline
anit-Stokes quality factor $Q_\text{aS}$ & $3.2 \times 10^7$ \\
\hline
pump quality factor $Q_\text{p}$ & $< 2.8 \times 10^7$ \\
\hline
$\Delta/2\pi$ & -20 ... +102 MHz \\
\hline
$g_0/2\pi$ & $(396.5 \pm 1.8) \ \text{Hz}$ \\
\hline
intracavity pump power $P_\text{cav}$ & 0 ... 126 W \\
\hline
$G/2\pi$ & 0 ... 39 MHz \\
\hline
\end{tabular}
  \label{tab:shape-functions}
\end{table}

\section{Conclusion and Outlook}
Using an optically doubly-resonant silica micro-rod-resonator, we have experimentally demonstrated Brillouin optomechanical strong coupling for high-frequency phonons (11 GHz) in the back-scattering regime. We observe normal-mode splitting and an avoided crossing in the optical emission spectrum, which give unambiguous signatures of the system operating in the strong coupling regime. This large optomechanical coupling rate was achieved by utilizing silica’s electrostriction and very low optical loss. We would like to highlight that our silica system does not suffer from two-photon absorption and strong surface-induced optical losses, which currently preclude many other micro-scale devices, primarily those fabricated from silicon, from entering the strong coupling regime. The Brillouin interaction utilized here additionally provides the advantage that the signal (anti-Stokes) photons are well separated from the pump field due the high mechanical frequency and being back-scattered from the pump field. To the best of our knowledge, this platform demonstrates optomechanical strong coupling with the highest mechanical frequency reported thus far.

The strong coupling performance achieved here can be even further improved via several near-, mid-, and longer-term routes. These include: (i) optimizing the choice of cavity mode pairs to simultaneously increase $g_0$ and reduce the power-dependent detuning; (ii) using higher optical quality factor resonances (quality factors of $10^9$ can be routinely fabricated that have an amplitude decay rate $\kappa/2\pi = 100 \ \text{kHz}$), which will allow lower input pump powers to be employed; (iii) using crystalline materials to reduce the mechanical damping rate \cite{Renninger:18, Hofer:10}; (iv) performing the experiments at cryogenic temperatures to reduce the thermal occupation and decoherence rate, which also provides the further advantage of reducing the mechanical damping rate \cite{Vacher:80, LeFloch:03}; and (v) exploring the use of resonators fabricated from other materials with a larger photoelastic coupling such as $\textrm{As}_2\textrm{S}_3$. We would also like to highlight at this point, that the heating due to intrinsic material absorption in silica for 10~W of intracavity power is expected to be $< 100$~mK when operating at a base temperature of 4~K.

The united favorable properties of this Brillouin optomechanical system provide a rich avenue to develop a suite of new technologies including classical and quantum information-processing applications, sensors, and even provide a path for coherent X-band microwave-to-optical conversion.
As highlighted above, operating at cryogenic temperatures and using crystalline materials will dramatically reduce the mechanical damping rate. At 4~K, the 11~GHz mechanical mode will have a mean thermal occupation of $\bar{n} \simeq 7.6$, and making the conservative assumption that the mechanical damping rate reduces to $\gamma_\text{m} \simeq 2$~MHz, the ratio of the mechanical decoherence rate to the mechanical frequency is $\bar{n} \gamma_\text{m} / \omega_\text{m} = \bar{n}/Q \simeq 10^{-3}$, which means there are approximately $10^3$ oscillations before decoherence becomes significant. With such reductions to the mechanical damping rate, the mechanical coherence length can exceed the resonator circumference, or equivalently, the mechanical free-spectral range can exceed the mechanical decay rate, i.e. $v/\pi D > \gamma_\text{m}/2\pi$, where $D$ is the resonator diameter. This parameter regime will be easily entered for resonators of similar size at cryogenic temperatures and the mechanical resonances will be resolved. Of the numerous applications and further studies that can be performed in this regime, we would like to highlight that this system can readily enter and explore the quantum-coherent-coupling regime where $G > \bar{n} \gamma_\text{m}$. The coupling rates achieved so far ($G/2\pi \sim$ 40~MHz), should be compared with our conservative estimate for the decoherence ($\bar{n} \gamma_\text{m} /2\pi \sim$~15~MHz). Operating in the quantum-coherent-coupling regime allows optomechanical state-swap to be performed. Moreover, the regime allows Rabi-like oscillations with non-classical optical input states, such as single-photon Fock states~\cite{Akram:10, Khalili:10, Verhagen:12}, to be observed, which is a key outstanding goal in the field. Achieving the strong coupling regime for this system paves the way to pursue this outlook and perform Brillouin-based quantum control of light and sound.

\section{Methods}
\subsection{Identifying cavity-mode pairs}
In order to find a pair of optical cavity modes that can provide Brillouin optomechanical strong coupling, we first find pairs that give low-threshold Brillouin lasing \cite{Tomes:09, Grudinin:09} via Stokes scattering where the higher frequency cavity mode is driven. (For the present silica microresonator, this corresponded to $\sim$ 1 mW input power.) We then employ a thermal lock to the lower-frequency mode of the pair and perform heterodyne measurements of the anti-Stokes-scattered light to characterize the system as described in the main text.

\subsection{Determining the intracavity power}
We determine the intracavity pump power via transmission measurements using the following procedure. For low to moderate intracavity pump powers (< 40 W), the pump-mode linewidth (damping rate) is power independent as there are no significant optical nonlinearities present, such as four-wave mixing. We then use our knowledge of the pump linewidth and the dimensions of the cavity to compute the finesse of the pump mode $F_\text{p} = \Delta \nu_\text{FSR} / \Delta \nu_\text{p} = c/(\pi n d \Delta \nu_\text{p})$. Here, $\Delta \nu_\text{FSR}$ is the cavity’s free spectral range, $\Delta \nu_\text{p}$ is the pump-mode linewidth, and $d$ is the microresonator diameter. We then compare the observed transmission when thermally locked $T$ to the minimum transmission when on-resonance $T_0$ to determine the detuning $\delta_\text{p}$  of the pump laser from resonance. This requires knowledge of the linewidth and cavity coupling conditions, i.e. being under- or over-coupled. Knowing the detuning, total linewidth, finesse, and external cavity coupling rate, we determine the intracavity power via $P_\text{cav} = F_\text{p}/\pi \cdot P_\text{in} \cdot (1-T)/(1-\sqrt{T_0})$. Note that this expression is valid for the over-coupled condition and the sign in the denominator flips for the under-coupled condition.

For higher powers (>40 W), the above procedure cannot be used due to optical nonlinearities becoming significant and increasing the pump-mode loss. (Note that the signal-mode loss is not increased by these nonlinear effects.) In this regime, we determine the intracavity power via the linear relationship between the detuning, as obtained from a fit to the heterodyne spectra, and the pump-mode intracavity power (cf. Fig. 1E). This method gives consistent results with the above method for the low power regime.

\section*{Funding Information}
This work was supported by the Engineering and Physical Sciences Research Council (EP/N014995/1 and EP/K034480/1), H2020 Marie Sklodowska-Curie Actions (MSCA) (748519, CoLiDR), National Physical Laboratory Strategic Research Programme, a H2020 European Research Council (ERC) grant 756966 (CounterLight), and an ERC Advanced Grant (MOQUACINO). LDB acknowledges support from EPSRC through the CDT for Applied Photonics. JS acknowledges support via a Royal Academy of Engineering Fellowship.

\section*{Acknowledgments}
We would like to thank L. Freisem, G. J. Milburn, C. Morrison, K. Mølmer, J. Nunn, J. Price, Y. Henry Wen, and J. Zhang for useful discussions.

%%%%%%%%%%%%%%%%%%%%%%%%%%%%%%%%%%%%%%%%%%%%%%%%%%%%
\clearpage
%%%%%%%%%%%%%%%%%%%%%%%%%%%%%%%%%%%%%%%%%%%%%%%%%%%%
\begin{center}\begin{LARGE}Supplementary information\end{LARGE}\end{center}

\section*{System Hamiltonian}
We use a simplified model to describe our experimental results that comprises two optical whispering gallery modes that couple via the high-frequency mechanical oscillation. The resonance frequencies of the two optical modes are separated by approximately the frequency of the mechanical resonance (see Fig.~1B of the main text), and the system is coherently driven at a frequency close to the lower frequency (pump) optical resonance. The main dynamics then take place between the higher frequency optical (anti-Stokes) mode and the mechanical resonance, while the pump mode field is approximated as a complex-number, which acts to enhance the interaction strength. Additionally, we account for the optical Kerr effect in silica, which leads to pump-power dependent resonance frequency shifts of the pump and anti-Stokes modes. More details of the mechanical eigenmode structure and the Brillouin phase-matching conditions are given in the final section of this supplementary.

%This model is justified by the fact that out of the multitude of optical and mechanical eigenmodes present in the microresonator system the contribution of a mode to the scattering rate is only significant if the detuning is comparable or small with respect to the damping rates in the system.

%\begin{figure}[htbp]
%\begin{center}
%\includegraphics[width=0.9\textwidth]{./res_structure.png}
%\caption{Caption goes here...}
%\end{center}
%\label{Fig:Modes}
%\end{figure}

The Hamiltonian of this interacting three-mode system is then

\begin{equation}
\begin{split}
\hat{H} = &\hbar g_0 \left( \hat{a}_\text{p} \hat{a}^\dag_\text{aS} \hat{b} + \hat{a}_\text{p}^\dag \hat{a}_\text{aS} \hat{b}^\dag \right) + \hbar  \omega_\text{p} \hat{a}_\text{p}^\dag \hat{a}_\text{p} + \hbar \omega_\text{aS} \hat{a}^\dag_\text{aS} \hat{a}_\text{aS} \\ &+ \hbar \omega_m \hat{b}^\dag \hat{b}  \ \  + \hat{H}_\text{drive} + \hat{H}_\text{SPM} + \hat{H}_\text{XPM} \  . 
\end{split}
\end{equation}
Here $\omega_\text{p}$, $\omega_\text{aS}$, and $\omega_\text{m}$ denote the angular frequencies of the pump, anti-Stokes, and mechanical modes; $\hat{a}_\text{p}$, $\hat{a}_\text{aS}$, and $\hat{b}$ are their field operators, respectively, and $g_0$ is the Brillouin optomechanical coupling strength. Throughout this supplementary material, hats denote operators, and tildes denote their Fourier transforms. The additional terms are given by \cite{Mandel:95}:

\begin{equation}
\begin{split}
\hat{H}_\text{drive} = \hbar \Omega^* e^{i\omega_\text{L} t} \hat{a}_\text{p} + \hbar \Omega e^{- i \omega_\text{L} t} \hat{a}^{\dag}_\text{p} \ \ \\ \ \text{with} \ \ \ \Omega = \sqrt{\kappa_\text{e} P_\text{in} / (\hbar \omega_\text{p})} ,
\end{split}
\end{equation}
which models the coherent drive. The remaining terms are the self phase modulation (SPM) and cross phase modulation (XPM):
\begin{equation}
\hat{H}_\text{SPM} = \frac{1}{2} \hbar \chi_\text{s} (\hat{a}^\dag_\text{p} \hat{a}_\text{p})^2 \, ,
\end{equation}
\begin{equation}
\hat{H}_\text{XPM} = 2 \hbar \chi_\text{x} (\hat{a}^\dag_\text{p} \hat{a}_\text{p})(\hat{a}^\dag_\text{aS} \hat{a}_\text{aS}) \, .
\end{equation}
Here, $\chi_\text{s}$ and $\chi_\text{x}$ are each linearly related to the $\chi^{(3)}$ third order susceptibility tensor.

\subsection*{Optical Kerr effect}
The third order nonlinearity present in silica leads to the optical Kerr effect, which is usually described in terms of a refractive index shift that linearly depends on the light intensity. The self- and cross-phase modulation terms lead to a frequency shift of the pump and anti-Stokes resonances depending on the circulating intensity in the pump mode. We can neglect self-phase modulation of the anti-Stokes mode since the power circulating in this mode is many orders of magnitude lower than the power in the coherently driven pump mode.

%Both, a high intensity light beam itself, and other modes (which may differ in frequency, polarisation) having a finite intensity overlap with the former beam show the optical Kerr effect and have a refractive index depending on the intensity of the former beam. Equivalently the action of the high intensity beam can be understood as a picked-up phase that depends on the intensity, and the corresponding effects are called self phase modulation (SPM) and cross phase modulation (XPM).

The pump-power dependent frequency shifts of the resonances are given by:
\begin{equation}
\omega_\text{p} = \omega_\text{p,0} - \chi \hat{n}_\text{p}
\end{equation}
\begin{equation}\label{eq:xpm_omega}
\omega_\text{aS} = \omega_\text{aS,0} - 2 \chi \hat{n}_\text{p}
\end{equation}
where the zero-subscripts indicate the initial frequencies, and we use the fact that in isotropic media (like fused silica) we have $\chi_\text{s} = \chi_\text{x} = \chi$, which assumes the pump and anti-Stokes modes have unity spatial overlap.

Equivalently, the resonance frequency shift can be expressed as due to an intensity dependent change in the refractive index:
\begin{equation}\label{eq:spm}
n(I_\text{p})=n_0+  n_2 I_\text{p}   \quad \text{self phase modulation}
\end{equation}
\begin{equation}\label{eq:xpm_n}
n(I_\text{p})=n_0+ 2 n_2 I_\text{p}   \quad \text{cross phase modulation}
\end{equation}
where $I_\text{p}$ denotes the pump intensity, and $n_2$ is the nonlinear refractive index, which is proportional to $\chi^{(3)}$.

The factor of 2 in Eqs~(\ref{eq:xpm_n} \& \ref{eq:xpm_omega}) is present if the modes have a perfect intensity overlap. The overlap is not unity in our experiment as the pump and anti-Stokes modes are not part of the same mode family and the precise value of the overlap is not known. Nevertheless, the mode overlap may be estimated from our experimental data, which is used to check the agreement between our model and the experiment. A more general treatment is given in the next section.

For the case where the optical modes overlap well, the refractive index change described by Eqs (\ref{eq:spm} \& \ref{eq:xpm_n}) leads to the detuning $\Delta = \omega_\textrm{aS} - \omega_\textrm{L} - \omega_\textrm{m} = \omega_\textrm{aS} - \omega_\textrm{p} - \omega_\textrm{m} - \delta_\textrm{p}$ changing with the pump power circulating in the cavity $P_\textrm{cav}$ in the following way:
\begin{equation}
\Delta - \Delta_0 = -  \frac{\omega_p}{n} \left( 2 n_2  \frac{P_\textrm{cav}}{A_\textrm{eff}} - n_2  \frac{P_\textrm{cav}}{A_\textrm{eff}} \right) \ \  ,
\end{equation}
where we used $\omega_\text{aS} \approx \omega_\text{p}$, and, $A_\text{eff}$ is the effective mode area of the pump mode. Thus, including the initial (low pump-power limit) detuning $\Delta_0$, the detuning then depends on the pump power as
\begin{equation}\label{eq:dynamicKerr}
\Delta \approx \Delta_0 - \frac{\omega_0 n_2}{n A_\textrm{eff}} \cdot P_\textrm{cav} \quad .
\end{equation}
It should be noted here that this expression is valid only when the pump and anti-Stokes modes overlap well. For smaller intensity overlaps and different mode shapes of the anti-Stokes mode, the slope of the linear relationship of Eq.~(\ref{eq:dynamicKerr}) can change and even change sign.

The thermal effect that is used to lock the pump optical resonance to the pump laser (see Opt. Express \textbf{12}, 4742 (2004)) is considered to have only a small effect on the detuning as the refractive index change due to temperature affects both the pump and anti-Stokes modes in the same way. Since the anti-Stokes mode is not perfectly overlapping with the pump mode whose absorption acts as the heat source, a significant difference of the thermo-optic responses of the two modes may also contribute to the power-dependent detuning observed in our experiment.

\subsection*{Intensity overlap in the optical Kerr effect}
The refractive index change via the optical Kerr effect that an optical mode A experiences under the presence of optical power in a (partly) overlapping mode B (cross phase modulation) is given via
\begin{equation}
\Delta n_A = 2 n_2 \frac{ \int I_B(\vec{x}) |u_A(\vec{x})|^2 d^3 x}{\int |u_A(\vec{x})|^2 d^3 x } \ \ , 
\end{equation}
where $u_A(\vec{x})$ is the mode field function of mode A, defined via $\vec{E}(\vec{x},t) = \vec{E}_0 e^{-i\omega t} u(\vec{x})$, and $I_B(\vec{x})$ is the intensity distribution of mode B, given by
\begin{equation}
I_B(\vec{x}) = E_B \frac{c}{n} \frac{|u_B(\vec{x})|^2}{\int |u_B(\vec{x})|^2 d^3 x} \ \ ,
\end{equation}
where $E_B$ is the energy stored in mode B. Putting this together, we obtain the expression containing the intensity overlap integral:
\begin{equation}
\Delta n_A = 2 n_2 E_B \frac{c}{n} \frac{ \int |u_A(\vec{x})|^2 |u_B(\vec{x})|^2 d^3 x}{\int |u_A(\vec{x})|^2 d^3 x \int |u_B(\vec{x}')|^2 d^3 x' }
\end{equation}
The frequency shift of the resonance frequency $\nu_A = N \Delta \nu_\text{FSR} = N c /(n_A L)$  is then simply $\Delta \nu_A / \nu_A = - \Delta n_A/n_A$. Here $N$ denotes the longitudinal mode number, and $L$ the round trip length.

Similarly the expression for self phase modulation is
\begin{equation}\label{eq:spmmodearea}
\Delta n = n_2 E \frac{c}{n} \frac{\int |u(\vec{x})|^4 d^3 x}{(\int |u(\vec{x})|^2 d^3 x)^2} \ \ .
\end{equation}

With $E = P \cdot t_\text{rt}$, with $t_\text{rt}$ the round trip time, and the effective mode area being defined as
\begin{equation}
A_\text{eff} = \frac{(\int |u(\vec{x})|^2 d^2 x)^2}{\int |u(\vec{x})|^4 d^2 x} \ \ 
\end{equation}
we see that Eq.~(\ref{eq:spmmodearea}) applied to the pump mode is equivalent to the expression $\Delta n = n_2 I_\text{p} = n_2 \cdot P_\text{cav}/A_\text{eff}$ used in the previous section.

Thus we obtain a more general expression for the contribution of the optical Kerr effect on the detuning
\begin{align}
\begin{split}
\Delta &= \Delta_0 - \frac{\omega_\text{p}}{n} (\Delta n_\text{aS} - \Delta n_\text{p}) \\
&= \Delta_0 - \frac{\omega_\text{p}}{n} \left( 2 n_2 E_\text{cav} \frac{c}{n} \frac{\int |u_\text{aS}|^2 |u_\text{p}|^2 d^3 x }{\int |u_\text{aS}|^2 d^3 x \int |u_\text{p}|^2 d^3 x} \right. \\ &\qquad \qquad \qquad \qquad \qquad \left. - n_2 E_\text{cav} \frac{c}{n} \frac{\int |u_\text{p}|^4 d^3 x}{(\int |u_\text{p}|^2 d^3 x)^2} \right) \\
&= \Delta_0 - \frac{\omega_\text{p} n_2 L P_\text{cav}}{n} \left( \frac{2 L \int |u_\text{aS}|^2 |u_\text{p}|^2 r dr d\theta}{ L^2 \int |u_\text{aS}|^2 r dr d\theta \cdot \int |u_\text{p}|^2 r dr d\theta} \right. \\ &\qquad \qquad \qquad \qquad \qquad \qquad \qquad  \left. - \frac{L \int |u_\text{p}|^4 r dr d\theta}{L^2 (\int |u_\text{p}|^2 r dr d\theta)^2} \right) \\
&= \Delta_0 - \frac{\omega_\text{p} n_2}{n \int |u_\text{p}|^2 r dr d\theta} \cdot P_\text{cav} \cdot \left( 2 \frac{\int |u_\text{aS}|^2 |u_\text{p}|^2 r dr d\theta}{\int |u_\text{aS}|^2 r dr d\theta} \right. \\ &\qquad \qquad \qquad \qquad \qquad \qquad \qquad \qquad  \left. - \frac{\int |u_\text{p}|^4 r dr d\theta}{\int |u_\text{p}|^2 r dr d\theta} \right)  , 
\end{split}
\end{align}
where we changed to polar coordinates, carried out the $\phi$-integration, and are left with an expression containing the difference of an overlap and an inverse effective mode area in the azimutal plane of the cylindrically symmetric microresonator.

This can be written as
\begin{equation}
\Delta = \Delta_0 - \frac{\omega_\text{p} n_2}{n A'} \cdot P_\text{cav}  \ \  ,
\end{equation}
where $A'$ relates to the mode overlap, has dimensions of an area, and is used in the main text. From the slope of the dependence of the detuning on the intracavity pump power (see Fig. 1 (E) of main text), we experimentally determine $A' = 3.9 \ \mu \text{m}^2$, based on the assumption that only the optical Kerr effect affects the detuning.

For comparison, the mode overlap integral associated with the Brillouin optomechanical coupling is given by
\begin{equation}
I_\text{om} = \int  u_\text{m} u_\text{aS} u_\text{p} r dr d\theta   \ \  ,
\end{equation}
where $u_\text{m}$ is the mechanical mode field function, 
compare \cite{Bahl:12}.

\subsection*{Simplifying the Hamiltonian}
In order to simplify the Hamiltonian, we perform a change of basis in Hilbert space, transforming to rotating frames with respect to the pump and anti-Stokes modes, and to a detuned frame with respect to the mechanical mode. The self- and cross-phase modulation we treat as a detuning that depends on the intracavity pump power, as described in the previous section. We also linearize the cubic Brillouin interaction treating the pump-mode field operator as a complex number. Starting from the Hamiltonian
\begin{equation}
\begin{split}
\frac{\hat{H}}{\hbar} = &g_0 \left( \hat{a}_\text{p} \hat{a}^\dag_\text{aS} \hat{b} + \hat{a}_\text{p}^\dag \hat{a}_\text{aS} \hat{b}^\dag \right) +  \omega_\text{p} \hat{a}_\text{p}^\dag \hat{a}_\text{p} \\ &+ \omega_\text{aS} \hat{a}^\dag_\text{aS} \hat{a}_\text{aS}  + \omega_m \hat{b}^\dag \hat{b}  \  ,
\end{split}
\end{equation}
we enter the rotating frame
\begin{equation}
\frac{\hat{H}}{\hbar} = g_0 (\hat{a}_\text{p} \hat{a}_\text{aS}^\dag \hat{b} + \hat{a}_\text{p}^\dag \hat{a}_\text{aS} \hat{b}^\dag) - \Delta \hat{b}^\dag \hat{b} \ \ ,
\end{equation}
and then make the approximation
\begin{equation}
\hat{a}_\text{p} \rightarrow \alpha = |\alpha| e^{i \phi} \qquad \hat{a}_\text{p}^\dag \rightarrow \alpha^{*} = |\alpha| e^{-i \phi} \ \ .
\end{equation}
Picking $\phi=0$, the Hamiltonian further simplifies to
\begin{equation}\label{eq:hamiltonian}
\frac{\hat{H}}{\hbar} = G (\hat{a}_\text{aS}^\dag \hat{b} + \hat{a}_\text{aS} \hat{b}^\dag) - \Delta \hat{b}^\dag \hat{b} \ , 
\end{equation}
where $G = g_0 |\alpha|$ is the intracavity-pump-power enhanced optomechanical coupling strength. At this point we put in the intracavity power dependence for $G$ and $\Delta$ and we arrive at the Hamiltonian used to model our experiment.

%In Eq.~(\ref{eq:hamiltonian}), both $G$ and $\Delta$ depend on the intracavity pump power. The dependence is given through Eq.~(\ref{eq:dynamicKerr}) for $\Delta$ and for the connection between $G$ and the pump power $P_\text{cav}$ we obtain 

It is useful to relate the intracavity pump amplitude $\alpha$ to the intracavity power via
\begin{equation}
|\alpha|^2 = \langle \hat{n} \rangle = \frac{P_\text{cav}}{ \hbar \omega_\text{p}} t_\text{rt}
\end{equation}
where $P_\text{cav} / \hbar \omega_\text{p} $ is the mean circulating pump photon rate and $t_\text{rt}$ the round trip time. We then have
\begin{equation}
G = g_0 |\alpha| =  g_0 \sqrt{\frac{\pi n d}{\hbar \omega_\text{p} c} P_\text{cav}} \ .
\end{equation}
Here, we inserted $t_\text{rt} = \pi n d / c$, where $d$ is the resonator diameter.

%%%%%%%%%%%%%%%%%%%%%%%%%%%%%%%%%%%%%%%%%%%%%%%%%%
\section*{Quantum Langevin equations of motion}
The open quantum system dynamics of the mode operators can be computed using the quantum Langevin equations
\begin{align}
\dot{\hat{a}}_\text{aS} &= - i \left[ \hat{a}_\text{aS}, \frac{\hat{H}}{\hbar}\right] - \kappa \hat{a}_\text{aS} + \sqrt{2 \kappa_i} \hat{a}_\text{in}^{(i)} + \sqrt{2 \kappa_e} \hat{a}_\text{in}^{(e)} \ , \\
\dot{\hat{b}} &= - i \left[ \hat{b}, \frac{\hat{H}}{\hbar} \right] - \gamma_m \hat{b} + \sqrt{2 \gamma_m} \hat{b}_\text{in} \ .
\end{align}
Here, $\gamma_m$ is the mechanical (amplitude) damping rate, and $\kappa$ is the total optical (amplitude) decay rate, which consists of intrinsic $\kappa_i$ and extrinsic $\kappa_e$ contributions, where $\kappa = \kappa_i + \kappa_e$. The two optical vacuum inputs and the mechanical thermal input noise terms are described by $\hat{a}_\text{in}^{(i)}$, $\hat{a}_\text{in}^{(e)}$, and $\hat{b}_\text{in}$, respectively. Substituting, $\hat{H}/\hbar = G (\hat{a}_\text{aS}^\dag \hat{b} + \hat{a}_\text{aS} \hat{b}^\dag) - \Delta \hat{b}^\dag \hat{b}$, we solve these coupled differential equations using the Fourier transform to obtain
\begin{align}
- i \omega \tilde{a}_\text{aS} &= - i G \tilde{b} - \kappa \tilde{a}_\text{aS} + \sqrt{2 \kappa_i} \tilde{a}_\text{in}^{(i)} + \sqrt{2 \kappa_e} \tilde{a}_\text{in}^{(e)} \\
- i \omega \tilde{b} &= - i (G \tilde{a}_\text{aS} - \Delta \tilde{b}) - \gamma_m \tilde{b} + \sqrt{2 \gamma_m} \tilde{b}_\text{in} \ \ .
\end{align}
Writing this as a matrix equation
\begin{equation}
\begin{gathered}
 \begin{pmatrix} \kappa - i \omega & i G \\ i G & \gamma_m - i (\omega + \Delta) \end{pmatrix} \begin{pmatrix} \tilde{a}_\text{aS} \\ \tilde{b} \end{pmatrix}
\\ =
  \begin{pmatrix}
 \sqrt{2 \kappa_i} \tilde{a}_\text{in}^{(i)} + \sqrt{2 \kappa_e} \tilde{a}_\text{in}^{(e)}  \\
\sqrt{2 \gamma_m} \tilde{b}_\text{in}
   \end{pmatrix} \ ,
\end{gathered}
\end{equation}
and solving for $\tilde{a}_\text{aS}$ and $\tilde{b}$, we obtain
\begin{gather}\label{eq:solution1}
  \begin{pmatrix} \tilde{a}_\text{aS} \\ \tilde{b} \end{pmatrix} = D \cdot \begin{pmatrix} - i (\omega + \Delta) + \gamma_m & -i G \\ - i G & - i \omega + \kappa \end{pmatrix} \ \times ... \\
\times \  \begin{pmatrix}
 \sqrt{2 \kappa_i} \tilde{a}_\text{in}^{(i)} + \sqrt{2 \kappa_e} \tilde{a}_\text{in}^{(e)}  \\
\sqrt{2 \gamma_m} \tilde{b}_\text{in}
   \end{pmatrix}
\end{gather}
where
\begin{equation}\label{eq:dee}
D = \frac{1}{(- i \omega + \kappa) (- i (\omega + \Delta) + \gamma_m) + G^2} \ \ .
\end{equation}
%In the following we will use $\hat{a}$ interchangeably with $\hat{a}_\text{aS}$ whenever considering the anti-Stokes optical mode.

\subsection*{Cavity input-output theory}
The field operator that describes the output of the cavity via the external decay channel introduced by the taper is obtained via
\begin{equation}
\tilde{a}_\text{out} = \tilde{a}_\text{in}^{(e)} - \sqrt{2 \kappa_e}  \tilde{a}_\text{aS} \ .
\end{equation}
We can account for extra loss in the detection system by applying a beam splitter transformation with transmissivity $t = \sqrt{\eta_\text{det}}$ and reflectivity $r = \sqrt{1 - \eta_\text{det}}$ to this output mode. Here, $\eta_\text{det}$ is the total detection efficiency after the out-coupling from the cavity into the tapered fiber, and we obtain
\begin{equation}
\tilde{a} = \sqrt{\eta_\text{det}} (  \tilde{a}_\text{in}^{(e)} - \sqrt{2 \kappa_e} \tilde{a}_\text{aS} ) + \sqrt{1 - \eta_\text{det}} \tilde{a}_\text{in}^{(bs)} \ ,
\end{equation}
where $\tilde{a}_\text{in}^{(bs)}$ represents the vacuum noise introduced by the open port of the beam splitter, which models the loss associated with the finite detection efficiency. Putting in the first line of Eq.~(\ref{eq:solution1}) we obtain
\begin{align}\label{eq:atilde}
\begin{split}
\tilde{a}(\omega)  &= 2 \sqrt{\eta_\text{det}} ( i G ) D \sqrt{ \kappa_e \gamma_m}   \ \cdot \ \tilde{b}_\text{in} \\
&\quad - 2 \sqrt{\eta_\text{det}} \sqrt{\kappa_e \kappa_i} D (-  i ( \omega + \Delta ) + \gamma_m ) \ \cdot \ \tilde{a}_\text{in}^{(i)} \\
&\quad + \sqrt{\eta_\text{det}} ( 1 - 2 \kappa_e D (-i (\omega + \Delta ) + \gamma_m ) ) \ \cdot \ \tilde{a}_\text{in}^{(e)} \\
&\quad + \sqrt{1 - \eta_\text{det}} \ \cdot \ \tilde{a}_\text{in}^{(bs)} \\
&= \ \ \  A_1(\omega) \tilde{b}_\text{in}(\omega) + A_2(\omega) \tilde{a}_\text{in}^{(i)}(\omega) \\
&\quad + A_3(\omega) \tilde{a}_\text{in}^{(e)}(\omega) + A_4(\omega) \tilde{a}_\text{in}^{(bs)}(\omega)
\end{split}
\end{align}
where in the last row we abbreviated the coefficients in front of the noise operators.
%%%%%%%%%%%%%%%%%%%%%%%%%%%%%%%%%%%%%%%%%%%%%%%%%%%
\section*{Heterodyne power spectrum}

In our experiment we measure the light backscattered from the cavity using heterodyne (more precisely: rotating homodyne) detection. Given the mode $\hat{a}(t)$ impinging on the detector, this rotating field quadrature is described by
\begin{equation}
\hat{X}_\theta (t) = \frac{1}{\sqrt{2}} \left( \hat{a}^\dag (t) e^{i \omega_\text{het} t} + \hat{a}(t) e^{-i \omega_\text{het} t}\right)
\end{equation}
where the heterodyne frequency is given by the difference between the frequency of the anti-Stokes cavity resonance and the freely chosen local oscillator frequency $\omega_\text{het} = \omega_\text{aS} - \omega_\text{LO}$.

%In the following we'll omit the hats on operators assuming it is clear from the context what is and what is not an operator.
%We can easily convince ourselves that the operator $X$ is hermitian $X^\dag (t) = X (t)$.

%To compare the predictions of our model to the observed spectra, we calculate the power spectrum of the rotating field quadrature $\hat{X}$, which our rotating homodyne detection scheme is continuously measuring.

We write the power spectral density in terms of the Fourier transformed quadrature operators
\begin{equation}\label{eq:sxxomega}
S_{XX}(\omega) = \int_{-\infty}^{\infty} d \omega' \langle \tilde{X}^\dag(\omega) \tilde{X}(\omega') \rangle \ .
\end{equation}
Firstly,
\begin{align}
\tilde{X}(\omega) &= \frac{1}{\sqrt{2}} \int d t e^{i \omega t} ( \hat{a}^\dag (t) e^{i \omega_\text{het} t} + \hat{a}(t) e^{-i \omega_\text{het} t} ) \\
&= \frac{1}{\sqrt{2}} ( \tilde{a}^\dag ( \omega + \omega_\text{het} )  + \tilde{a} ( \omega - \omega_\text{het} )  )  \ . 
\end{align}
Re-inserting this expression, we obtain for the power spectrum
\begin{align}
\begin{split}
\label{eq:sxxas}
S_{XX} ( \omega) = \frac{1}{2} \int d \omega' ( \langle ( \tilde{a} ( -\omega - \omega_\text{het}))^\dag \tilde{a}(\omega' - \omega_\text{het}) \rangle \\ + \langle \tilde{a} ( \omega - \omega_\text{het} ) ( \tilde{a} (  - \omega' - \omega_\text{het} ) )^\dag \rangle ) \ ,
\end{split}
\end{align}
where the two other cross terms are zero for a mechanical thermal state. In order to compute $S_{XX}(\omega)$, we take the expression for $\tilde{a}(\omega)$ from Eq.~(\ref{eq:atilde}) and compute its adjoint
\begin{align}
\label{eq:daggeredas}
\begin{split}
(\tilde{a}(\omega))^\dag &= A_1^*(\omega)( \tilde{b}_\text{in}(\omega))^\dag + A_2^*(\omega)( \tilde{a}_\text{in}^{(i)}(\omega))^\dag  \\ &\qquad + A_3^*(\omega)( \tilde{a}_\text{in}^{(e)}(\omega))^\dag  + A_4^*(\omega)( \tilde{a}_\text{in}^{(bs)}(\omega))^\dag \\
&=  A_1^*(\omega) \tilde{b}_\text{in}(-\omega) + A_2^* ( \omega) \tilde{a}_\text{in}^{(i)}(-\omega) \\ &\qquad + A_3^*(\omega) \tilde{a}_\text{in}^{(e)}(-\omega) + A_4^*(\omega) \tilde{a}_\text{in}^{(bs)}(-\omega)
\end{split}
\end{align}
Inserting this into Eq.~(\ref{eq:sxxas}) and assuming the noise terms are delta correlated we obtain
\begin{multline}
S_{XX}(\omega) = \frac{1}{2} \left[ | A_1 (-\omega - \omega_\text{het} ) |^2 n_b  \right. \\ +  | A_1 (\omega - \omega_\text{het} ) |^2 (n_b + 1) + |A_2 (\omega - \omega_\text{het})|^2 \\ \left. + |A_3 (\omega - \omega_\text{het})|^2 + |A_4 (\omega - \omega_\text{het})|^2 \right] \ .
\end{multline}
Then, inserting the $A$-terms from Eqs~(\ref{eq:atilde} \& \ref{eq:daggeredas}) gives
\begin{align}
\begin{split}
S_{XX}(\omega) &= \frac{1}{2} \ + \ 2 \eta_\text{det} \kappa_e \gamma_m G^2 n_b \ \times ... \\ &\qquad \times \ \left( |D(-\omega - \omega_\text{het})|^2 + |D(\omega - \omega_\text{het})|^2 \right) \ .
\end{split}
\end{align}
Lastly, putting in the expression for $D(\omega)$ from Eq.~(\ref{eq:dee}) we find the power spectrum given by Eq~(\ref{eq:longExpression}).
\begin{figure*}[h!tbp!]
\begin{align}
\label{eq:longExpression}
\begin{split}
&S_{XX}(\omega) = \frac{1}{2} \ + \  2 \eta_\text{det} \kappa_e \gamma_m G^2 n_b \times \\ &\times \left( \frac{1}{(G^2 - (\omega + \omega_\text{het})(\omega + \omega_\text{het} - \Delta) + \kappa \gamma_m)^2 + ((\omega + \omega_\text{het}) \gamma_m + (\omega + \omega_\text{het} - \Delta) \kappa )^2 } \right.  \\ 
&+ \left. \frac{1}{(G^2 - (\omega - \omega_\text{het})(\omega - \omega_\text{het} + \Delta) + \kappa \gamma_m)^2 + ((\omega - \omega_\text{het}) \gamma_m + (\omega - \omega_\text{het} + \Delta) \kappa )^2 } \right) \ .
\end{split}
\end{align}
\end{figure*}
%

%%%%%%%%%%%%%%%%%%%%%%%%%%%%%%%%%%%%%%%%%%%%%%%%%%%
\section*{Normal-mode splitting and strong coupling}

If the Brillouin optomechanical coupling between the anti-Stokes and mechanical modes is sufficiently strong then normal-mode splitting and an avoided crossing will be observed in the spectra. This phenomenon is associated with the emergence of hybrid optical-mechanical modes, which is critical for many classical and quantum applications as described in the main text.

In order to find the eigenfrequencies and damping rates of the system we diagonalize a non-Hermitian Hamiltonian that includes the damping terms. These terms are chosen such that in the Heisenberg equation of motion they explicitly produce the same damping terms as found in the Langevin equations of motion.

The non-Hermitian Hamiltonian for our system in the rotating frame reads

\begin{equation}
\frac{\hat{H}}{\hbar} = G (\hat{a}_\text{aS}^\dag \hat{b} + \hat{a}_\text{aS} \hat{b}^\dag) - (\Delta + i \gamma_m) \hat{b}^\dag \hat{b} \ - \ i \kappa \hat{a}_\text{aS}^\dag \hat{a}_\text{aS} \ \  .
\end{equation}
We can write this expression as a matrix equation
\begin{gather}\label{eq:solution}
  \frac{\hat{H}}{\hbar} =  \begin{pmatrix} \hat{a}_\text{aS}^\dag \ ,& \hat{b}^\dag \end{pmatrix}  \cdot \begin{pmatrix}  - i \kappa & G \\ G & - (\Delta + i\gamma_m) \end{pmatrix}
  \begin{pmatrix}
\hat{a}_\text{aS} \\  \hat{b}
   \end{pmatrix} \ \ .
\end{gather}
Eigenfrequencies and effective damping rates of the system follow from this matrix. We write down the characteristic polynomial and find its roots:

\begin{equation}
\lambda^2 + \left( \Delta + i(\kappa + \gamma_\text{m})\right) \lambda + i \kappa \Delta - \kappa \gamma_\text{m} - G^2 = 0
\end{equation}

The two roots, and thus the complex eigenvalues of the matrix, are
\begin{align}
\begin{split}
\lambda_\pm &= - \frac{\Delta}{2} - i \frac{\kappa + \gamma_m}{2} \\ &\qquad \pm \sqrt{G^2 + \left(\frac{\Delta}{2}\right)^2 - \left( \frac{\kappa - \gamma}{2} \right)^2 - i \Delta \frac{\kappa - \gamma_m}{2}} \ \ .
\end{split}
\end{align}

We know that the complex square root of $z = x + i y$ is given by
\begin{equation}
\sqrt{z} = \sqrt{ \frac{ | z | + x}{2}} + i \ \text{sgn}^{+}(y) \cdot \sqrt{\frac{|z| - x}{2}}
\end{equation}
with $x = G^2 + \left(\frac{\Delta}{2}\right)^2 - \left( \frac{\gamma_\text{m} - \kappa}{2} \right)^2$ and $y = \frac{\gamma_\text{m} - \kappa}{2} \Delta$.

From the negative of the real parts of this pair of eigenvalues we obtain the eigenfrequencies, and from the negative of the imaginary parts we obtain the damping rates associated with the modes.
%\begin{align}
%\omega_\pm &= \Delta/2 \pm \sqrt{\frac{|z|+x}{2}} \\
%\gamma_\pm &= \frac{\kappa + \gamma_\text{m}}{2} \mp \text{sgn}^{+}(y) \cdot \sqrt{\frac{|z| - x}{2}}
%\end{align}
The general case of $\Delta \neq 0$ can not be brought into a simpler form. To discuss the transition from weak to strong coupling we discuss the case of zero detuning, which contains the essential features.

For $\Delta = 0$ the complex eigenvalues are:
\begin{equation}\label{eq:delta0}
\lambda_{\pm, \Delta = 0} = - \frac{\kappa + \gamma_\text{m}}{2} i \pm \sqrt{G^2 - \left(\frac{\kappa - \gamma_\text{m}}{2}\right)^2}
\end{equation}
We see that the real part of this equation is $0$ as long as $G < |\kappa - \gamma_\text{m}|/2$, corresponding to the case of degenerate eigenfrequencies of optical and mechanical modes in the rotating frame. In the case of vanishing coupling $G \rightarrow 0$, we observe that the imaginary parts approach the uncoupled damping rates of the modes, i.e. $- \text{Im}(\lambda_{\pm, \Delta = 0})|_{G \rightarrow 0} = \kappa , \gamma$.

As the coupling between the modes increases, the damping is redistributed among the new eigenstates up until the point where $G$ becomes larger than $|(\kappa - \gamma_\text{m})/2|$, at which point the term under the square root becomes positive and the damping rates of both eigenmodes are equal, given by $(\kappa + \gamma_\text{m})/2$. It is also here that normal mode splitting occurs, as the real part of Eq.~(\ref{eq:delta0}) divides into the branches of the square root. In this parameter regime we have hybrid optical-mechanical modes with equal contributions from both of the oscillators.

We observe that normal-mode splitting formally occurs as soon as $G > |(\kappa - \gamma_\text{m})/2|$, but that this normal mode splitting is not resolved until the coupling strength $G$ also overcomes the effective damping rate of the hybrid optical-mechanical modes $(\kappa + \gamma_\text{m})/2$. 

The criterion of strong coupling is fulfilled if the coupling is strong enough so that the normal mode splitting  at $\Delta = 0$ can actually be resolved. This occurs if 
\begin{equation} 
G > \frac{\kappa + \gamma_m}{2} \  \  
\end{equation}    
i.e. when G becomes larger than the damping rate of the hybrid optical-mechanical modes.

\begin{figure}[thbp]
\centering
\includegraphics[width=0.73\columnwidth]{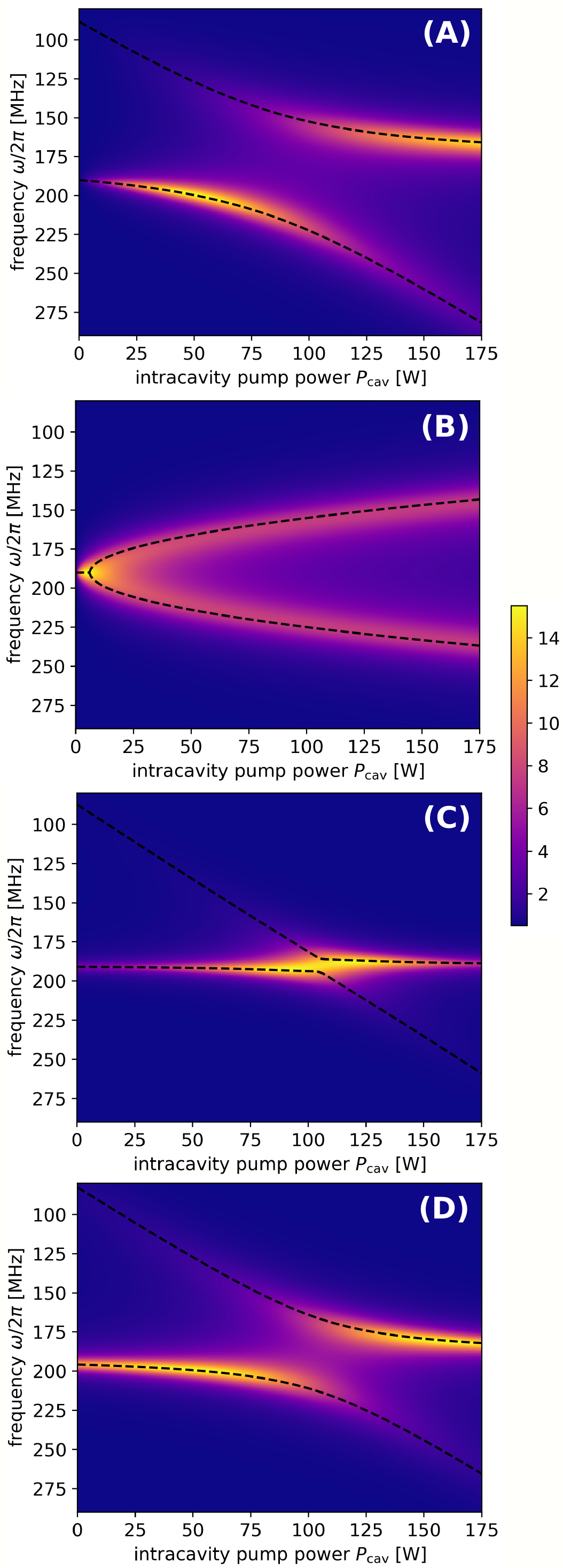}
\caption{Contour plots of optical spectra and superimposed eigenfrequencies (black dashed line) for different forms of the power dependence of the detuning $\Delta$ and optomechanical coupling strength $G$. A value of $1/2$ on the colour scale corresponds to optical vacuum. (A) Spectrum and eigenvalues for the parameters implemented in the experiment described in the main text. (B) Hypothetical situation with the same parameters however the detuning is constant at $0$ (optical Kerr effect switched off). (C) and (D) Hypothetical scenarios where the Kerr effect is switched on, but the optomechanical coupling strength $G$ is constant, at 10 MHz (C), and 25 MHz (D). Here we see that the spectra and eigenvalues are symmetric around the zero detuning point found just above 100 W. }\label{fig:contourplots}
\end{figure}

Figure \ref{fig:contourplots} plots the predicted heterodyne spectra as a function of the intracavity power to highlight different physical scenarios of variable detuning and coupling strength.

%\begin{figure}[!htbp]
%  \centering
%  \begin{minipage}[t]{0.49\textwidth}
%    \includegraphics[width=\textwidth]{contour+eigenvalues.png}
%    \caption{Countour plot and eigenfrequencies vs coupling strength $G$ for hypothetical constant $0$ detuning.}
%  \end{minipage}
%  \hfill
%  \begin{minipage}[t]{0.49\textwidth}
%    \includegraphics[width=\textwidth]{contour+eigenvalues_w_Kerr.png}
%    \caption{Countour plot and eigenfrequencies vs coupling strength $G$ for dynamical detuning, starting at $101.8$ MHz.}
%  \end{minipage}
%\end{figure}

\section*{Brillouin phase matching}
In Brillouin scattering processes both energy and quasi-momentum are conserved. With Brillouin Stokes scattering, where a pump photon is converted into a lower frequency Stokes photon and a phonon, the phase-matching conditions are $\hbar \omega_\text{p} = \hbar \omega_\text{S} + \hbar \omega_\text{m}$ and $\hbar k_\text{p} = \hbar k_\text{S} + \hbar k_\text{m}$, where the indices p, S and m stand for pump, Stokes and mechanical frequencies and wave vectors, respectively. In anti-Stokes scattering where a pump photon and a phonon combine to produce a bluer anti-Stokes photon, we have $\hbar \omega_\text{p} + \hbar \omega_\text{m} = \hbar \omega_\text{aS}$ and $\hbar k_\text{p} + \hbar k_\text{m} = \hbar k_\text{aS}$. 

\begin{figure}[!thbp]
\centering
\includegraphics[width=0.82\columnwidth]{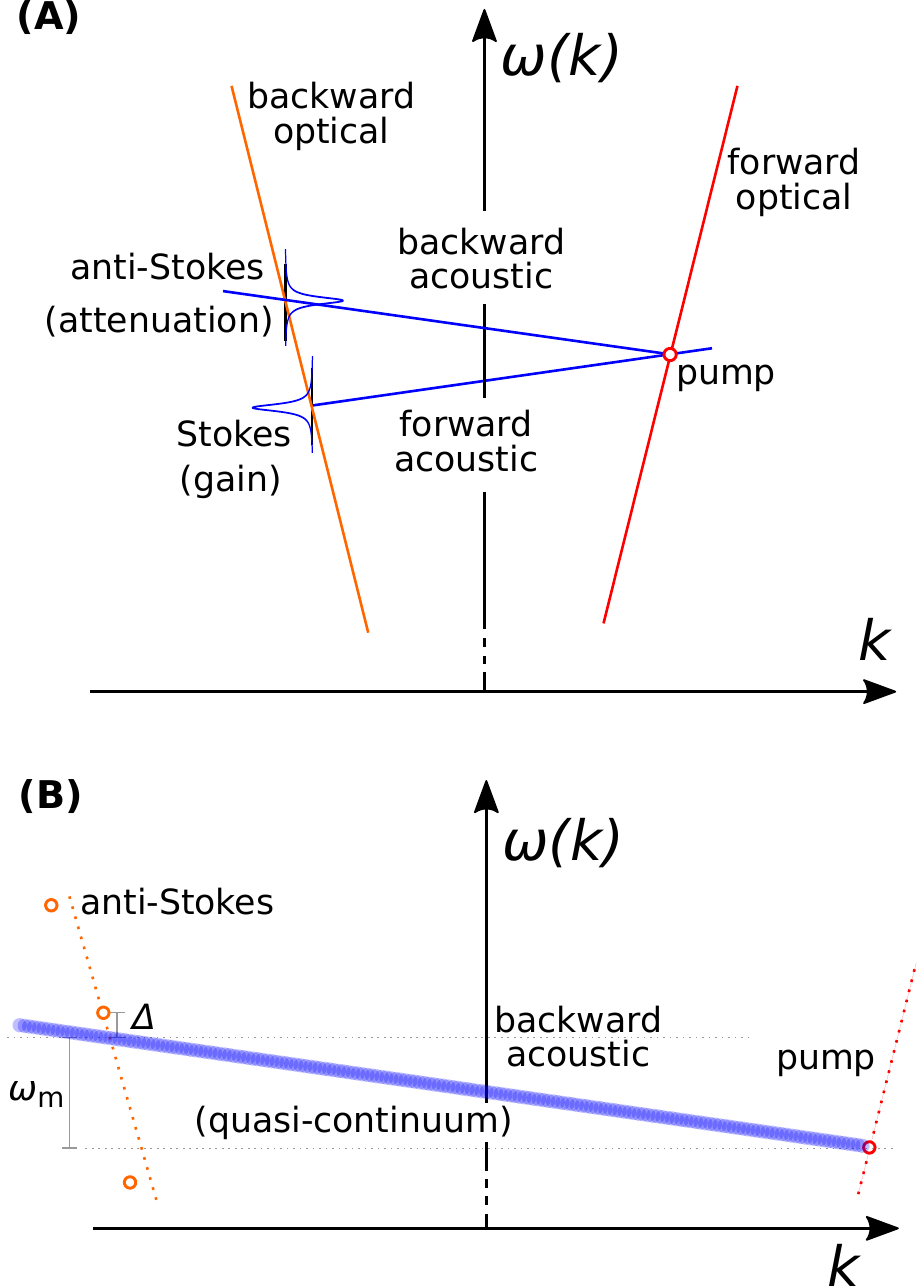}
\caption{Brillouin phase matching. (A) Continuous optical and acoustic dispersion relation for backward Brillouin scattering in a waveguide. (B) Discrete optical and acoustic modes by periodic boundary conditions in a cavity (as in this work). The acoustic dispersion relation is a quasi-continuum. Mechanical frequency $\omega_\text{m}$ and detuning $\Delta$ are indicated (compare Fig 1, main text).}\label{fig:BrillPhaseMatch}
\end{figure}

In figure \ref{fig:BrillPhaseMatch} the phase matching condition is depicted in terms of the dispersion relations of light and sound, for a waveguide system (A) and for a cavity (B). In the waveguide, there is a continuum of both optical and mechanical modes available. Picking a particular pump frequency (red circle) there are two 'automatically' phase matched back-scattering processes, corresponding to the intersection points of the acoustic dispersion relation (blue) for forward and backward going sound waves with the optical dispersion relations (red and orange). These two conditions are: (i) Stokes scattering from the forward acoustic mode, associated with gain and stimulated-Brillouin scattering, and (ii) anti-Stokes scattering from the counter-propagating acoustic mode, associated with attenuation and cooling. 

When considering a travelling wave resonator as in this work, the dispersion relations are no longer continuous, but discrete, as indicated by the circles in figure \ref{fig:BrillPhaseMatch} (B). Different transverse optical mode families will experience different dispersion and the dots representing them will lie on different dotted lines, of which an example is given for one anti-Stokes mode family in the figure. As the acoustic damping rate is larger than the frequency separation between adjacent acoustic frequency-eigenmodes of the cavity (mechanical free spectral range), the acoustic dispersion forms a quasi-continuum of overlapping frequency eigenmodes. In such a resonator, due to the discrete nature of the optical cavity modes, the phase matching is no longer 'automatic', and an optomechanical detuning $\Delta$ was introduced to describe the small frequency mismatch between the mechanical frequency and the optical frequency difference (see figure \ref{fig:BrillPhaseMatch}).
Our simplified model detailed here does not explicitly detail this mechanical quasi-continuum mode structure. Specifically, the mechanical field operator $\hat{b}$ in the model then describes the linear combination of mechanical eigenfrequency components that contribute to the phase-matching in the narrow frequency band as selected by the pair of optical cavity resonances.

\end{document}